\documentclass[a4paper,11pt]{article}
\pdfoutput=1 % if your are submitting a pdflatex (i.e. if you have
\usepackage{jheppub} % for details on the use of the package, please
\usepackage{bm}
\usepackage[T1]{fontenc} % if needed
\newtheorem{theorem}{Theorem}[section]
\newtheorem{lemma}[theorem]{Lemma}

\title{\boldmath Generalizations of the Pontryagin and Husain-Kucha\v{r} actions to manifolds with boundary}

\author[a,d]{J. Fernando Barbero G.,}
\author[a]{Bogar D\'{\i}az,}
\author[b,d,e]{Juan Margalef-Bentabol}
\author[c,d]{and Eduardo J. S. Villase\~nor\,}

\affiliation[a]{Instituto de Estructura de la Materia, CSIC. Serrano 123, 28006 Madrid, Spain}
\affiliation[b]{Laboratory of Geometry and Dynamical Systems, Department of Mathematics, EPSEB, Universitat Polit\`ecnica de Catalunya, BGSMath, Barcelona, Spain}
\affiliation[c]{Universidad Carlos III de Madrid. Avda.\  de la Universidad 30, 28911 Legan\'es, Spain}
\affiliation[d]{Grupo de Teor\'{\i}as de Campos y F\'{\i}sica Estad\'{\i}stica. Instituto Gregorio Mill\'an (UC3M). Unidad Asociada al Instituto de Estructura de la Materia, CSIC, Madrid, Spain}
\affiliation[e]{Institute for Gravitation and the Cosmos \& Physics Department, Penn State, University Park, PA 16802, USA}

\emailAdd{fbarbero@iem.cfmac.csic.es}
\emailAdd{bdiaz@iem.cfmac.csic.es}
\emailAdd{juanmargalef@ucm.es}
\emailAdd{ejsanche@math.uc3m.es}

\abstract{In this paper we study a family of generalizations of the Pontryagin and Husain-Kucha\v{r}  actions on manifolds with boundary. In some cases, they describe well-known models---either at the boundary or in the bulk---such as 3-dimensional Euclidean general relativity with a cosmological constant or the Husain-Kucha\v{r} model. We will use Hamiltonian methods in order to disentangle the physical and dynamical content of the systems that we discuss here. This will be done by relying on a geometric implementation of the Dirac algorithm in the presence of boundaries recently proposed by the authors.}

\arxivnumber{1906.09820}

\begin{document}
\maketitle
%\flushbottom

\section{Introduction}

The Pontryagin action for an $SO(3)$-connection on a 4-dimensional manifold induces a Chern-Simons theory on its boundary because the Pontryagin Lagrangian (written in terms of the curvature as $F_i\wedge F^i$) is the exterior differential of the Chern-Simons 3-form \cite{Chern}.

A generalization of the Pontryagin action with an interesting geometrical interpretation was discussed in \cite{FB1}. The main idea of that paper was to take \emph{two} $SO(3)$-connections $A_-^i$ and $A_+^i$ as dynamical variables and consider the Lagrangian $F_{+i}\wedge F_-^i$ (where $F^i_\pm$ denote the curvatures of $A^i_\pm$ respectively) on a 4-manifold without boundary. Notice that, by coupling the curvatures in this way, the internal symmetry corresponds to a single copy of $SO(3)$. The dynamics defined by this generalized action was interpreted in \cite{FB1} by showing that it can be written as the action of the model discussed by Husain and Kucha\v{r} in \cite{HK} (referred to in the following as the HK model). This leads to an immediate physical interpretation of the dynamics that can be conveniently studied by looking at the Hamiltonian formulation. In fact, once a foliation of the spacetime manifold is introduced, the phase space of the system becomes the cotangent bundle of a configuration space consisting of $SO(3)$-connections. The constraints of the theory generate ``internal'' $SO(3)$ rotations and 3-dimensional diffeomorphisms, hence, the model describes 3-geometries. This phase space is exactly the one employed in the Ashtekar formulation of general relativity (GR), which differs from the HK model in the presence of an additional scalar constraint that crucially generates the full dynamics of GR.

The main purpose of the present paper is to study some generalizations of the Pontryagin and Husain-Kucha\v{r} actions to 4-dimensional manifolds \emph{with boundary}. A large family of them lead to the HK model in the bulk and interesting dynamics at the boundary, including---among other models selected by specific choices of the coupling constants---3-dimensional Euclidean GR with an \emph{arbitrary} cosmological constant.

The layout of the paper is the following. After this introduction, we study  in section \ref{sec_2connHK} the two-connection action derived from the Lagrangian $F_{+i}\wedge F_-^i$ on a manifold \emph{with boundary}. By introducing an affine combination of the two connections and using also their difference, we relate this action to the standard one for the HK model. Along the way, we find out that a particular choice of the parameter in the aforementioned combination leads to the HK model in the bulk and 3-dimensional Euclidean GR at the boundary with a particular value for the cosmological constant. Section \ref{sec_genHK} is devoted to discussing a five-parameter family of generalizations of the Pontryangin and HK actions to manifolds with boundary suggested by the results of the preceding section. As we will show, the dynamics at the spacetime boundary defined by some of them describes 3-dimensional Euclidean GR with an \emph{arbitrary} cosmological constant. The Hamiltonian formulation of the models proposed in the paper is derived and discussed in section \ref{sec_HamHK}. We will also study how the physical degrees of freedom can be parametrized in a way that may open the possibility of having a concrete and useful description of the reduced phase space. We end with our conclusions and some comments in section \ref{sec_Conclusions}.

We use a mixed notation in the paper: we will avoid spacetime indices but will use Penrose's abstract internal $SO(3)$ indices $i,j,k\ldots$ These will be raised and lowered with the help of the invariant $SO(3)$ metric $\delta_{ij}$. Finally, we denote the internal $SO(3)$ volume form as $\epsilon_{ijk}$.

\section{The two-connection formulation of the Husain-Kucha\texorpdfstring{\v{r}}{r} model in manifolds with boundary}\label{sec_2connHK}

Let us consider the following generalization of the Pontryagin action to a 4-dimensional manifold $M$ with boundary $\partial M$
\begin{equation}\label{action12}
  S(A_+,A_-):=\int_M F_{+i}\wedge F_-^i =S(A_-,A_+)\,,
\end{equation}
where $F_{\pm}^i$ are the curvatures of $A_\pm^i$
\begin{equation}\label{curvatures}
F_\pm^i:=\mathrm{d}A_\pm^i+\frac{1}{2}\epsilon^i_{\ jk}A_\pm^j\wedge A_\pm^k\,.
\end{equation}
The stationarity of the action \eqref{action12} gives the following conditions
\begin{subequations}
\begin{align}
   D_-F_+^i&=0\,,& D_+F_-^i&=0\,, \label{2con_bulk}\\
   \jmath^*F_+^i&=0 \,, &\jmath^*F_-^i&=0\,, \label{2con_bound}
\end{align}
\end{subequations}
where $D_\pm$ denote the covariant derivatives defined by the connections $A^i_\pm\,$, the map $\jmath:\partial M\hookrightarrow M$ is the natural inclusion, and $\jmath^*$ denotes the pullback to the boundary. As we can see, the field equations in the bulk have the form of ``interleaved Bianchi identities'' whereas at $\partial M$ we have ``natural boundary conditions'' telling us that the pullbacks of both connections must be flat. By making use of the Bianchi identities, it is possible to write \eqref{2con_bulk} in a way that shows that these equations are, in fact, of first order
\begin{equation}\label{2con_bulk_Bianchi}
  \epsilon^i_{\,\,jk}(A_+^j-A_-^j)\wedge F_+^k=0\,,\qquad\qquad \epsilon^i_{\,\,jk}(A_+^j-A_-^j)\wedge F_-^k=0\,.
\end{equation}
By relying on \eqref{2con_bulk_Bianchi}, it is straightforward to show the existence of non-trivial solutions to \eqref{2con_bulk} and \eqref{2con_bound} by taking $A_+^i=A_-^i$ such that the pullback of $F_{\pm}^i$ to $\partial M$ is zero. Another possibility is taking two flat connections in $M$.

In order to see the relation between the usual formulation of the HK model and its two-connection formulation when $\partial M$ is not necessarily empty, we define \cite{FB1}
\begin{subequations}\label{alpha12}
\begin{align}
e^i&:= {A_+^i}-{A_-^i}\,,\\
A^i&:= \alpha A_-^i+(1-\alpha) A_+^i= A_+^i+ \alpha (A_-^i-A_+^i)= A_+^i- \alpha e^i\,,
\end{align}
\end{subequations}
where $\alpha$ is a real constant. A direct computation now gives
\begin{align}\label{Lb}\begin{split}
F_{\!-\,i}\wedge F_+^i=& \frac{(1-\alpha)^2+\alpha^2}{2} \epsilon_{ijk} e^i\wedge e^j \wedge F^k + \frac{\alpha\left( 1-\alpha\right) \left( 1-2\alpha\right)}{2} \epsilon_{ijk} D e^i \wedge e^j \wedge e^k \\
+& \alpha \left( \alpha-1 \right) D e_i \wedge D e^{i}+\left( 2\alpha-1\right)     D e_i \wedge F^i + F_i \wedge F^i\,,\end{split}
\end{align}
where $D$ denotes the covariant derivative defined by $A^i$. Among the terms appearing in the preceding expression we recognize the Pontryagin and HK Lagrangians. In addition, we have all the other objects that can be naturally written in terms of the $A^i$ and $e^i$ fields.

The field equations obtained by varying the action with respect to $e^i$ and $A^i$ can be written as
\begin{subequations}
\begin{align}
 \epsilon_{ijk}e^j\wedge F^k=0\,,\qquad\qquad \epsilon_{ijk}De^j\wedge e^k&=0\,,\label{field_equations_bulk}\\
 \jmath^*\left( De^i-\frac{1}{2}(1-2\alpha)\epsilon^i_{\,\,jk}e^j\wedge e^k\right)&=0\,.\label{field_equations_bound2}\\
 \jmath^* \left(F^i-\frac{1}{2}\alpha(\alpha-1)\epsilon^i_{\,\,jk}e^j\wedge e^k\right)&=0\,,\label{field_equations_bound1}
\end{align}
\end{subequations}
As we can see, in addition to the usual equations for the HK model \eqref{field_equations_bulk}, we also get the equations \eqref{field_equations_bound2} and \eqref{field_equations_bound1} on $\partial M$. These can be interpreted as equations for the pullbacks of $A^i$ and $e^i$ to the spacetime boundary. Their solutions can be mapped in a one-to-one way to those of the 3-dimensional Euclidean Einstein equations with a particular value of the cosmological constant. An easy way to see this is by noting that the choice $\alpha=1/2$ leads to
\begin{equation}
\jmath^* \left(De^i \right)=0\,,\qquad\qquad \jmath^*\left(F_i+\frac{1}{8}\epsilon_{ijk} e^j \wedge e^k\right)=0 \,, \label{b3d}
\end{equation}
i.e. the field equations for Euclidean 3-dimensional GR on $\partial M$, written in terms of a connection and a triad field, with cosmological constant $\Lambda=1/4$. This is a remarkable result because it shows an interesting relation between 3-dimensional GR in $\partial M$ and the HK model in $M$. Notice that, although the physical interpretation of the preceding equations in terms of $A^i$ and $e^i$ is clear, some solutions to them are easier to describe in the two-connection formulation, in particular the one mentioned above in terms of flat connections.

\section{Generalizations of the Pontryagin and Husain-Kucha\v{r} actions }\label{sec_genHK}

The previous result suggests that it may be possible to find actions related to the ones discussed above describing 3-dimensional GR with an arbitrary cosmological constant---and more general models---on the boundary. Indeed, let us introduce the following five-parameter generalization of the action defined by the Lagrangian \eqref{Lb} in a manifold with boundary:
\begin{align}\label{hkpbcc}\begin{split}
S_{\bm \alpha}(e,A):=\int_M &\left( \alpha_1 \epsilon_{ijk} e^i \wedge e^j \wedge F^k+ \alpha_2 De_i \wedge De^i +\alpha_3  F_i \wedge F^i \right.\\
&\hspace*{4cm}\left.+\alpha_4 \epsilon _{ijk}D e^i \wedge e^j \wedge e^k +\alpha_5 F_i\wedge D e^i
\right) \,.\end{split}
\end{align}
Of course, this can be written in terms of two connections $A_\pm^i$ using \eqref{alpha12}, however, the interpretation of the field equations will be more transparent in terms of the usual variables $e^i$ and $A_i$ so we will use them from now on. The equations of motion are
\begin{subequations}\label{EcsGen}
\begin{equation}\label{HKbulk}
 \hspace*{24mm}\left( \alpha_1-\alpha_2 \right) \epsilon_{ijk}  e^j \wedge F^k=0\,,\hspace*{2.1cm} \left( \alpha_1-\alpha_2 \right) \epsilon_{ijk} De^j \wedge e^k=0 \,,\,\,\,\,
\end{equation}
\begin{equation}\label{BE}
 \jmath^* \!\left(2\alpha_2 De_i+\alpha_4 \epsilon_{ijk} e^j \wedge e^k\!+\!\alpha_5 F_i\right)    =0 \,,\,
 \jmath^* \!\left(\alpha_1\epsilon_{ijk}e^j\wedge e^k\!+\!2\alpha_3F_i+\alpha_5 D e_i\right)=0 \,.
\end{equation}
\end{subequations}

\noindent Notice that the boundary equations \eqref{BE} can be written as before in terms of the pullbacks of $A^i$ and $e^i$  to the boundary.

Particular choices of the constants ${\bm \alpha}=(\alpha_1 , \alpha_2, \alpha_3, \alpha_4, \alpha_5)$ lead to several interesting models. For instance, if $\alpha_1\neq\alpha_2$ the equations in the bulk are the ones corresponding to the standard HK model. On the other hand, if $\alpha_1=\alpha_2$ the action can be written as an integral over the boundary $\partial M$ and corresponds to the Mielke--Baekler model \cite{Mielke, MBH}
\begin{align}\label{Mielke-Baeckler}\begin{split}
&S_{MB}(\jmath_\partial^*e,\jmath_\partial^*A)=\int_{\partial M}\jmath_\partial^*\Big(\alpha_1e_i\wedge De^i+\alpha_3 (A_i\wedge \mathrm{d}A^i+\frac{1}{3}\epsilon_{ijk}A^i\wedge A^j \wedge A^k)\\
&\hspace*{8cm}+\frac{\alpha_4}{3}\epsilon_{ijk}e^i\wedge e^j \wedge e^k+\alpha_5 F_i\wedge e^i\Big)\,.\end{split}
\end{align}
This explains why we do not have any equations in the bulk in this case, in full analogy with the behaviour of the Pontryagin action.

If $4\alpha_2\alpha_3- \alpha_5^2=0$, the equations on the boundary \eqref{BE} are degenerate: they imply degenerate metrics or become an underdetermined system of first-order partial differential equations for the field variables.

When $4\alpha_2\alpha_3- \alpha_5^2\neq0$, the equations \eqref{BE} can be written as
\begin{subequations} \label{BE2}
\begin{align}
\jmath^* \left(  F_i \right) &= \jmath^* \left( - \frac{2\alpha_1 \alpha_2-\alpha_4\alpha_5}{4\alpha_2 \alpha_3 -\alpha_5^2} \epsilon_{ijk} e^j \wedge e^k  \right)   \,,\\
 \jmath^* \left( D e_i \right)&= \jmath^* \left(- \frac{2\alpha_3 \alpha_4-\alpha_1\alpha_5}{ 4\alpha_2 \alpha_3 -\alpha_5^2} \epsilon_{ijk} e^j \wedge e^k \right)\,.
\end{align}
\end{subequations}
They describe 3-dimensional Euclidean gravity with torsion. Furthermore, if $2\alpha_3 \alpha_4-\alpha_1 \alpha_5=0$ we have standard 3-dimensional Euclidean GR with a cosmological constant equal to
\[\displaystyle \frac{4\alpha_1\alpha_2-2\alpha_4\alpha_5}{4\alpha_2\alpha_3- \alpha_5^2}.\]
Alternatively, if $2\alpha_1\alpha_2-\alpha_4\alpha_5=0$ we have the teleparallel description of 3-dimensional Euclidean GR with cosmological constant \cite{Kawai,BV1,BV2} equal to
\[
-\left(\frac{2\alpha_3 \alpha_4-\alpha_1\alpha_5}{ 4\alpha_2 \alpha_3 -\alpha_5^2} \right)^2\,.
\]

Although our methods can be applied with no difficulty for all possible values of the $\bm \alpha$ parameters (and, in fact, are especially effective for this purpose), in the following we will restrict ourselves to the generic case $\alpha_1\neq\alpha_2$ and $4\alpha_2\alpha_3- \alpha_5^2  \neq 0$ for three reasons:
\begin{itemize}
\item As the analysis of the previous section shows, there are solutions to the full set of field equations that are compatible in the bulk and the boundary.
\item The equations at the boundary admit natural and interesting interpretations.
\item This sector has received some attention in recent years  \cite{Cacciatori, BGMR,BC}. Even the more restricted case $2\alpha_3 \alpha_4-\alpha_1 \alpha_5=0$, has been related to the presence of an Immirzi-like parameter in three dimensions \cite{Livine, Basu}.
\end{itemize}
In addition to the conditions on the $\bm \alpha$ parameters mentioned before, in the following we will also impose a non-degeneracy condition for the $e^i$.

\section{Hamiltonian formulation for the generalized models}\label{sec_HamHK}

The Hamiltonian description of the dynamics of the models presented in section \ref{sec_genHK} provides a useful way to disentangle their meaning. For instance, it gives information about the compatibility of the dynamical equations in the interior of $M$ and in the boundary $\partial M$. The Dirac algorithm can, in fact, be thought of as a way to obtain conditions (constraints) that must be imposed on the configuration variables and their conjugate momenta to have consistent dynamics (here we will follow \cite{Dirac_geom}, similar information can be obtained by using the GNH method \cite{Gotay2,BPV,tesisJuan}).

The constraints restrict the possible initial data for the field equations. As we will see, in the present case some of them are associated with the boundary and others with the bulk. Both types are not independent but are expected to have a non-trivial interplay. In the case of familiar boundary conditions (say of the Dirichlet type for elliptic PDE's) we know that they can be freely chosen so that solutions of the equations---with the required uniqueness and regularity---can then be found. In other instances (such as free boundary problems in elasticity), the character of the possible conditions/dynamical equations on the boundary and their interplay with the dynamics in the bulk is different. In the present example a thorough discussion of the existence and uniqueness of solutions to the full set of constraints is certainly necessary but lies beyond the reach of the Hamiltonian methods used in the paper. In any case, in order to find meaningful characterizations of the physical degrees of freedom, it is useful to find explicit parametrizations of the phase space submanifolds defined by the constraints so we will analyze this question in subsection \ref{solving}.

Finally, and for the sake of completeness, it is important to remember that the \textit{integrability} of the Hamiltonian vector fields is a separate non-trivial issue that must also be carefully considered.

In this section, we derive the Hamiltonian formulation corresponding to the generalized model defined by the action \eqref{hkpbcc} under the generic conditions on the parameters $\bm \alpha$ and the triads expelled in section \ref{sec_genHK}.

Let us consider 4-manifolds of the form $M=\mathbb{R}\times\Sigma$ where $\Sigma$ is a 3-dimensional manifold (possibly with boundary). The Lagrangian defined by the action \eqref{hkpbcc} after performing a 3+1 decomposition is
\begin{align*}\begin{split}
L({\mathrm{v}} )\!=\!\int_{\Sigma} &\Big( \!\left( v_A^{i}-D A^i_{\perp} \right)\!\wedge\!\left( \alpha_1  \epsilon_{ijk} e^{j} \!\wedge\! e^{k}+2\alpha_3 F_i +\alpha_5 D e_i\right)\!+ 2\epsilon_{ijk} e^i_{\perp} e^{j}\!\wedge\! \left(\alpha_1F^{k}+\alpha_4 D e^k \right)  \\
&\hspace*{2.6cm}+ \left( v_e^{i}+\epsilon^i{}_{jk} A^j_{\perp} e^{k}-D e^i_{\perp} \right) \!\wedge\! \left(2\alpha_2 D e_i+\alpha_4 \epsilon_{ilm}e^l\!\wedge e^m +\alpha_5 F_i\right)\Big)\,,\end{split}
\end{align*}
where the variables $A^i$ and $e^i$ are now an $SO(3)$-connection and a frame field on $\Sigma$ respectively, the fields $A_\perp^i$ and $e_\perp^i$ are $\mathfrak{so}(3)$-valued scalars on $\Sigma$ originating in the transverse components of the 4-dimensional fields with respect to the spacetime foliation, and
\[
\mathrm{v}:=((A_\perp^i,A^i,e_\perp^i,e^i),(v_{A_\perp}^i,v_A^i,v_{e_\perp}^i,v_e^i))
\]
denotes a point of the tangent bundle $T\mathcal{Q}$  of   the configuration space $\mathcal{Q}$ (defined by the transverse and tangent parts of the connections and ``frames'').  In this section we denote the covariant derivative defined by the connection $A^i$ as $D$ and the curvature as
\[
F_i:=\mathrm{d}A_i+\frac{1}{2}\epsilon_{ijk}A^j\wedge A^k\,.
\]
If we take $\mathrm{v}, \mathrm{w}$ in the same fiber of $T\mathcal{Q}$,
\begin{align*}
  \mathrm{v}& :=((A_\perp^i,A^i,e_\perp^i,e^i),(v_{A_\perp}^i,v_A^i,v_{e_\perp}^i,v_e^i))\,,\\
  \mathrm{w}& :=((A_\perp^i,A^i,e_\perp^i,e^i),(w_{A_\perp}^i,w_A^i,w_{e_\perp}^i,w_e^i))\,,
\end{align*}
we obtain the fiber derivative
\begin{align*}
&\left\langle F\!L \left(\mathrm{v}\right)|\mathrm{w}\right\rangle=  \int_{\Sigma} \left( w_A^{i}\wedge\big(  \alpha_1  \epsilon_{ijk} e^{j} \wedge e^{k}+2\alpha_3 F_i +\alpha_5 D e_i \big)\right. \hspace*{3cm}\\
& \hspace*{8cm}\left.+ w_e^{i} \wedge   \big(2\alpha_2 D e_i+\alpha_4 \epsilon_{ijk}e^j\wedge e^k +\alpha_5 F_i\big)  \right)\,.
\end{align*}
This implies that we have the following primary constraints
\begin{subequations}\label{consHKpb}
\begin{align}
{\bf C}_{\perp i}(\cdot)&:= {\bf P}_{\!\!\perp i}(\cdot) = 0 \,, & \hspace*{-3mm} {\bf C}_{i}(\cdot)&:=  \displaystyle {\bf P}_{i}(\cdot)-\!\!\int_{\Sigma} \!\cdot \wedge\left(  \alpha_1  \epsilon_{ijk} e^{j} \wedge e^{k}+2\alpha_3 F_i +\alpha_5 D e_i \right)=0\,,\\
{\bf c}_{\perp i}(\cdot)&:= {\bf p}_{\perp i}(\cdot) = 0     \,, &  \hspace*{-3mm} {\bf c}_{i}(\cdot)&:= {\bf p}_{i}(\cdot)- \!\!\int_{\Sigma} \! \cdot \wedge  \left(2\alpha_2 D e_i+\alpha_4 \epsilon_{ijk}e^j\wedge e^k +\alpha_5 F_i\right) = 0\,,
\end{align}
\end{subequations}
where here and in the following the points $(q,{\bf p})\in T^*\mathcal{Q}$ will be denoted as
\begin{equation*}
 (q,{\bf p}):=\big((A_\perp^i,A^i,e_\perp^i,e^i),({\bf P}_{\!\!\perp i},{\bf P}_{ i},{\bf p}_{\perp i},{\bf p}_{i})\big)\,.
\end{equation*}
Notice that ${\bf C}_{\perp i}(\cdot)$ and ${\bf c}_{\perp i}(\cdot)$ are linear functionals acting on $\mathfrak{so}(3)$-valued scalar functions on $\Sigma$ whereas ${\bf C}_{i}(\cdot)$ and ${\bf c}_{i}(\cdot)$ are linear functionals acting on $\mathfrak{so}(3)$-valued 1-forms.

The Hamiltonian is only defined on the primary constraint submanifold. A suitable extension of it to the full phase space of the model can be written as
\begin{align*}
H=\int_{\Sigma} & \left( D A^i_{\perp}\wedge \left( \alpha_1\epsilon_{ijk} e^{j}\wedge e^{k}+2\alpha_3 F_i +\alpha_5 D e_i\right)- 2\epsilon_{ijk}e^i_{\perp} e^{j}\wedge\left( \alpha_1 F^{k}+\alpha_4 De^k\right) \right.\\
&\hspace*{3.5cm}\left. - \left( \epsilon^i{}_{jk} A^j_{\perp} e^{k}- D e^i_{\perp} \right) \wedge \left(2\alpha_2 D e_i+\alpha_4 \epsilon_{ilm}e^l\wedge e^m +\alpha_5 F_i\right) \right)\,.
\end{align*}
As we can see, it is independent of the canonical momenta. This fact will have a reflection in the form of the Hamiltonian vector fields. By writing tangent vectors $Z\in T_{(q,{\bf p})}T^*\mathcal{Q}$ as
\begin{align*}
  Z&:=\left((q,{\bf p}),(Z^i_{A_\perp},Z^i_{A},Z^i_{e_\perp},Z^i_{e}, {\bf Z}_{{\bf P}_{\!\!\perp} i},{\bf Z}_{{\bf P} i}, {\bf Z}_{{\bf p}_{\perp }i},{\bf Z}_{{\bf p} i})\right)\,,
\end{align*}
the canonical symplectic form $\Omega$ acting on vector fields on $T^*\mathcal{Q}$ is
\begin{align*}
\Omega(X,Y)=&{\bf Y}_{{\bf P}_{\!\!\perp } i}\left(X^i_{A_\perp}\right)-{\bf X}_{{\bf P}_{\!\!\perp }i}\left(Y^i_{A_\perp}\right)+{\bf Y}_{{\bf P}i}\left(X^i_{A}\right)-{\bf X}_{{\bf P} i}\left(Y^i_{A}\right)\nonumber\\
+&{\bf Y}_{{\bf p}_{\perp} i}\left(X^i_{e_\perp}\right)\,-{\bf X}_{{\bf p}_{\perp} i}\left(Y^i_{e_\perp}\right)\,+{\bf Y}_{{\bf p} i}\left(X^i_{e}\right)\,-{\bf X}_{{\bf p} i}\left(Y^i_{e}\right)\,.
\end{align*}
The implementation of the geometric form of the Dirac algorithm described in \cite{Dirac_geom} is now a straightforward exercise. The main step is solving for the Hamiltonian vector field $X$ in the equation
\begin{align*}
\Omega(X,Y)&= \, {\bf {d}} H \left(Y\right) +\langle \lambda^i_{\perp} |{\bf {d}}{\bf C}_{\perp i}\rangle \left(Y\right)+\langle \lambda^i |{\bf {d}}{\bf C}_{i}\rangle \left(Y\right)+  \langle  \mu^i_\perp| {\bf {d}} {\bf c}_{\perp i} \rangle \left(Y\right)+ \langle \mu^i |{\bf {d}}{\bf c}_{i} \rangle \left(Y\right)\,,
\end{align*}
for every vector field $Y$. Here ${\bf {d}}$ denotes the exterior differential in phase space, $\langle\, \cdot\, |\, \cdot\, \rangle$ is the usual pairing, and the $\lambda_\perp^i$, $\lambda^i$,  $\mu_\perp^i$, $\mu^i$ are Dirac multipliers.

The final result of the analysis can be summarized in the conditions defining the constraint submanifold for the system and the specific form of the Hamiltonian vector fields. The constraints are
\begin{subequations}\label{constraints}
\begin{align}
  &\hspace*{-2mm}{\bf P}_{\!\!\perp i}(\cdot)=0\,,  &  \hspace*{-15mm}{\bf P}_{i}(\cdot)-\!\!\int_\Sigma \!\cdot\wedge\!\big(\alpha_1\epsilon_{ijk}e^j\wedge e^k+2\alpha_3F_i +\alpha_5 D e_i\big)\!=\!0\,,\label{c1}\\
  &\hspace*{-2mm}{\bf p}_{\perp i}(\cdot)=0\,,  &  \hspace*{-15mm}{\bf p}_{i}(\cdot)- \!\!\int_{\Sigma}  \!\cdot \wedge\!  \left(2\alpha_2 D e_i+\alpha_4 \epsilon_{ijk}e^j\wedge e^k +\alpha_5 F_i\right) \!=\!0\,,\label{c2}\\
  &\hspace*{-2mm}(\alpha_1-\alpha_2)\epsilon_{ijk}e^j\wedge F^k=0\,, &  (\alpha_1-\alpha_2)\epsilon_{ijk}De^j\wedge e^k\!=\!0\,,\label{c3}\\
  &\hspace*{-2mm}\imath_\partial^*\left(\!F^i\!-\!\frac{2\alpha_1\alpha_2-\alpha_4\alpha_5}{\alpha_5^2-4\alpha_2\alpha_3}\epsilon_{ijk}e^j\!\wedge e^k\right)=0\,, &  \imath_\partial^*\left(\!De^i\!-\!\frac{2\alpha_3\alpha_4-\alpha_1\alpha_5}{\alpha_5^2-4\alpha_2\alpha_3}\epsilon_{ijk}e^j\!\wedge e^k\right)\!=\!0\,,\label{c4}\\
  &\hspace*{-2mm}\imath_\partial^*\left(\!V_A^i-2\frac{2\alpha_1\alpha_2-\alpha_4\alpha_5}{\alpha_5^2-4\alpha_2\alpha_3}\epsilon_{ijk}\,e^j_\perp e^k\right)=0\,, &  \imath_\partial^*\left(\!V_e^i-2\frac{2\alpha_3\alpha_4-\alpha_1\alpha_5}{\alpha_5^2-4\alpha_2\alpha_3}\epsilon_{ijk}\,e^j_\perp e^k\right)\!=\!0\,,\label{c5}
\end{align}
\end{subequations}
where $\imath_\partial:\partial\Sigma\hookrightarrow\Sigma$ is the natural inclusion and the $\mathfrak{so}(3)$-valued 1-forms $V_A^k$ and $V_e^k$ satisfy the conditions
\begin{subequations}\label{conditions}
  \begin{align}
    \epsilon_{ijk}e^j\wedge V_A^k&=\epsilon_{ijk}e_\perp^jF^k\,,\label{conditions1}\\
    \epsilon_{ijk}e^j\wedge V_e^k&=\epsilon_{ijk}e_\perp^jDe^k\,.\label{conditions2}
  \end{align}
\end{subequations}
Here, the secondary constraints \eqref{c3} (on $\Sigma$) and \eqref{c4} (on $\partial\Sigma$) are obtained by requiring that the Hamiltonian vector field be tangent to the submanifold defined by the primary constraints \eqref{c1} and \eqref{c2}.  In addition to these secondary constraints, the tangency requirement gives conditions for the Dirac multipliers $\lambda^i$ and $\mu^i$. These are, again, of two types: conditions on $\Sigma$ and conditions on $\partial\Sigma$. The constraints \eqref{c5} are obtained by requiring the consistency of the values of the pullbacks of $\lambda^i$ and $\mu^i$ obtained by solving both types of conditions (notice that the conditions in the bulk would not be present if $\alpha_1=\alpha_2$). The values of $\lambda^i$ and $\mu^i$ are
\begin{align*}
  & \mu^i=De_\perp^i+\epsilon^i{}_{jk}A_\perp^ke^j+V_e^i\,, \\
  & \lambda^i=DA_\perp^i+V_A^i\,.
\end{align*}
The conditions \eqref{c5} define several sectors on the constraint submanifold. For instance, a straightforward computation shows that whenever
\begin{equation}\label{sector}
\imath_\partial^* \big(\epsilon_{ijk}e^i_\perp e^j\wedge e^k\big)=0\,,
\end{equation}
the constraints \eqref{c5} are satisfied. In the following we will concentrate on this particular sector because it has a clean and interesting interpretation that will be discussed in subsection \ref{gauge_symmetry}. The tangency condition on \eqref{sector} only fixes the transverse component of $\imath_\partial^*\mu^i_\perp$. The remaining components of $\imath_\partial^*\mu^i_\perp$, $\mu^i_\perp$ and $\lambda^i_\perp$ remain undetermined, with the only restriction that the bulk values of $\mu^i_\perp$ must be compatible with those at the boundary.

The Hamiltonian vector field is given by
\begin{subequations}\label{VectorField}
  \begin{align}
    &\hspace*{-3mm}X_{A_\perp}^i\!=\!\lambda^i_\perp\,, &X_{A}^i&\!=DA_\perp^i+V_A^i\,,\\
    &\hspace*{-3mm}X_{e_\perp}^i\!=\!\mu^i_\perp\,, & X^i_{e}&\!=De_\perp^i-\epsilon^i{}_{jk}A_\perp^je^k+V_e^i\,,\\
    &\hspace*{-3mm}{\bf X}_{{\bf P}_{\!\!\perp} i}(\cdot)\!=\!0\,,&\hspace*{-1mm}{\bf X}_{{\bf P} i}(\cdot)&\!=\!\!\!\int_\Sigma\!\!\cdot\wedge\!\left(2\alpha_1\epsilon_{ijk}X_e^j\wedge e^k\!+\!2\alpha_3 DX_{\!Ai}\!+\!\alpha_5\left( DX_{ei}\!+\!\epsilon_{ijk}X_A^j\wedge e^k \right)\right),\\
    &\hspace*{-3mm}{\bf X}_{{\bf p}_{\perp} i}(\cdot)\!=\!0\,,&\hspace*{-1mm}{\bf X}_{{\bf p} i}(\cdot)&\!=\!\!\!\int_\Sigma\!\!\cdot\wedge\!\left( 2\alpha_2\left(DX_{ei}\!+\!\epsilon_{ijk}X_A^j\wedge e^k\right)\!+\! 2\alpha_4\epsilon_{ijk}X_e^j\wedge e^k\!+\!\alpha_5 DX_{Ai}    \right),
  \end{align}
\end{subequations}

The dynamics on the boundary of the pullbacks of the fields $A_\perp$, $e_\perp$, $A^i$ and $e^i$ is given by the pullback to the boundary of the components of the Hamiltonian vector field,
\begin{subequations}\label{pb}
\begin{align}
\imath_\partial^* X^i_{A_{\perp}}        &=  \imath_\partial^* \lambda^i_{ \perp}  \,,&
\imath_\partial^* X^i_{A}    &=    \imath_\partial^*\left(D A^i_{\perp} -2\frac{2\alpha_1\alpha_2-\alpha_4\alpha_5}{4\alpha_2\alpha_3-\alpha_5^2}\epsilon^i{}_{jk} e^j_{\perp} e^k\right) \,,\\
\imath_\partial^* X^i_{e_{\perp}}        &=  \imath_\partial^*\mu^i_{\perp}   \,,&
\imath_\partial^* X^i_{e}  &=  \imath_\partial^*  \left(D e^i_{\perp} - \epsilon^i{}_{jk} A^j_{\perp} e^k-2\frac{2\alpha_3\alpha_4-\alpha_1\alpha_5}{4\alpha_2\alpha_3-\alpha_5^2}\epsilon^i{}_{jk} e^j_{\perp} e^k\right) \,.
\end{align}
\end{subequations}
It is interesting to note at this point that $X^i_{e_{\perp}}$, $X^i_{e}$, $X^i_{A_{\perp}}$ and $X^i_{A}$ are independent of the momenta and, hence, the dynamics of the system can be obtained without having to consider the remaining components of $X$. This is a consequence of the fact that the Hamiltonian is only defined at the primary constraint submanifold and can be extended to the full phase space in many ways. The extension that we have chosen is such that the Hamiltonian depends only on the configuration variables. In the traditional presentation of the Hamiltonian formulation of the HK model \cite{HK}, the canonical momenta are taken to be densitized triads so the preceding result may seem strange. It should be noted, however, that in that context the second class constraints \eqref{c1} and \eqref{c2} are solved (i.e.\ all the geometric objects, including the symplectic form, are pulled-back to the phase space submanifold defined by them) and, also, one is using (more or less implicitly) some kind of duality to represent canonical momenta as geometric objects of the same type as the configuration variables. Although the geometric representation of the dynamical objects that we are using here is different, both procedures are compatible. The main reason to follow the present one is because it is better suited to address functional analytic issues if deemed necessary. It is important to keep in mind that other approaches may be better suited for quantization.

We discuss now the meaning of the dynamical system defined by the integral curves of the Hamiltonian vector field given by \eqref{VectorField} and \eqref{pb} and some issues related to the parametrization of the constraint submanifold.

\subsection{Dynamics and gauge symmetries}\label{gauge_symmetry}

The previous Hamiltonian vector field, in particular the components $X^i_{e_{\perp}}$, $X^i_{e}$, $X^i_{A_{\perp}}$ and $X^i_{A}$, admits a clean interpretation on the constraint submanifold of the system. To this end, we will make use of the following useful result whose simple proof can be found in the appendix of \cite{barbero}:

\begin{lemma}\label{prop}
Let $\Sigma$ be an orientable 3-dimensional manifold. For given $w_i\in\Omega^2(\Sigma,\mathfrak{so}(3)^*)$ and $e^i\in\Omega^1(\Sigma,\mathfrak{so}(3))$ defining a volume form ${\bm \omega}:=\frac{1}{3!}\epsilon_{ijk}e^i\wedge e^j\wedge e^k$, consider the following system of equations in the unknowns $v^i\in\Omega^1(\Sigma,\mathfrak{so}(3))$
\begin{equation}\label{equations}
  \epsilon_{ijk}e^j\wedge v^k=w_i\,.
\end{equation}
Then the solution is
\begin{equation}\label{solution}
v^i=\frac{e^j\wedge w_j}{2{\bm \omega}} e^i- \frac{e^i\wedge w_j}{{\bm \omega}} e^j\,.
\end{equation}
Here and in the following $\displaystyle \frac{\bm\eta}{\bm \omega}$ denotes the function $\varphi$ satisfying ${\bm\eta}=\varphi {\bm \omega}$.
\end{lemma}

With the help of lemma \ref{prop} it is straightforward to see from eqn. \eqref{conditions1} that
\[
V_A^i=-\epsilon_{jk\ell}\left( e_\perp^j \frac{e^k\wedge F^\ell}{2{\bm \omega}}e^i+ e_\perp^k \frac{e^i\wedge F^\ell}{{\bm \omega}} e^j\right)
\]
which, on the constraint submanifold, can be written as
\[
V_A^i=-\epsilon_{jk\ell}e_\perp^k \frac{e^\ell\wedge F^i}{\bm \omega} e^j\,.
\]
In an analogous way, from eqn. \eqref{conditions2} we obtain on the constraint submanifold
\[
V_e^i=-\epsilon_{jk\ell}e_\perp^k\frac{e^\ell\wedge De^i}{\bm \omega} e^j\,.
\]
Using the expressions for $V_A^i$ and $V_e^i$, the equations for the integral curves of the Hamiltonian vector field \eqref{VectorField} for initial data on the constraint submanifold give
\begin{subequations}
\begin{align}
  \dot{A}^i & =D\tau^i-\epsilon_{jk\ell}\rho^k \frac{e^\ell\wedge F^i}{\bm \omega} e^j=D\tau^i+\imath_{\bm\rho}F^i\,,\label{Adot}\\
  \dot{e}^i & =D\rho^i-\epsilon^i_{\,\,jk}\tau^je^k-\epsilon_{jk\ell}\rho^k \frac{e^\ell\wedge De^i}{\bm \omega} e^j=D\rho^i-\epsilon^i_{\,\,jk}\tau^je^k+\imath_{\bm\rho}De^i\,,\label{edot}
\end{align}
\end{subequations}
where $\rho^i$ and $\tau^i$ are arbitrary functions of time (because the evolution of the $A_\perp^i$ and $e_\perp^i$ is arbitrary) subject to the condition that the vector field ${\bm \rho}$ defined by $\imath_{\bm \rho} e^i=\rho^i$ is tangent to the boundary of $\Sigma$ as a consequence of the constraint \eqref{sector}. In order to interpret \eqref{Adot} and \eqref{edot} it is convenient to take into account that, by using Cartan's formula, the Lie derivative of $A^i$ and $e^i$ along ${\bm \rho}$ can be written as
\begin{align*}
  &\pounds_{\bm \rho} A^i=D\left( \imath_{{\bm \rho}} A^i \right) +\imath_{{\bm \rho}} F^i\,,\\
  &\pounds_{\bm \rho} e^i\,\,= D \left( \imath_{{\bm \rho}} e^i \right) + \imath_{{\bm \rho}} \left( D e^i\right) + \epsilon^i{}_{jk} e^j\big( \imath_{{\bm \rho}} A^k \big)\,.
\end{align*}
Combining these expressions with \eqref{Adot} and \eqref{edot} we immediately get
\begin{subequations}
\begin{align}
  \dot{A}^i & =\pounds_{{\bm \rho}} A^i+D(\tau^i-\imath_{\bm \rho} A^i)\,,\label{Adot2}\\
  \dot{e}^i & =\pounds_{\bm \rho} e^i-\epsilon^i{}_{jk}(\tau^j-\imath_{\bm \rho} A^j)e^k\,.\label{edot2}
\end{align}
\end{subequations}
The interpretation of the dynamics of the model in the bulk is clear from \eqref{Adot2} and \eqref{edot2}: it is a combination of spatial diffeomorphisms and internal $SO(3)$ rotations for initial data satisfying the constraints \eqref{c3}. This is, of course, the known physical interpretation of the dynamics of the HK model. This interpretation can be extended to the manifold $\Sigma$, including its boundary, because the $\bm \rho$ vector field is tangent to $\partial \Sigma$.

The dynamics at the boundary of the pulled-back fields can be read from \eqref{pb}. In particular, denoting pullbacks to the boundary of $\Sigma$ with a $\partial$ subindex we have
\begin{subequations}
\begin{align}
 \dot{A}_\partial^i & =D_\partial\tau_\partial^i-2\frac{2\alpha_1\alpha_2-\alpha_4\alpha_5}{4\alpha_2\alpha_3-\alpha_5^2}\epsilon^i_{\,jk}\rho_\partial ^je_\partial ^k\,,\label{Adot2bound}\\
 \dot{e}^i_\partial & =D_\partial\rho_\partial^i-\epsilon^i_{\,jk}\tau_\partial ^je_\partial ^k
 -2\frac{2\alpha_3\alpha_4-\alpha_1\alpha_5}{4\alpha_2\alpha_3-\alpha_5^2}\epsilon^i{}_{jk} \rho^j_{\partial} e^k_\partial
 \,.\label{edot2bound}
\end{align}
\end{subequations}
Some comments are in order now: To begin with, the initial data on the boundary must satisfy the constraints \eqref{c4} and, furthermore, the condition \eqref{sector}. The parameters $\tau_\partial^i$ are arbitrary whereas, the $\rho_\partial^i$ are not all independent as a consequence of the constraint \eqref{sector}. If $\tau_\partial^i$ and $\rho_\partial^i$ were completely arbitrary the equations \eqref{Adot2bound} and \eqref{edot2bound} would correspond exactly to the dynamics of Euclidean GR with a cosmological constant and torsion \cite{BGMR}, defined on the spacetime boundary. However, the condition \eqref{sector} allows only two of the components of $\rho_\partial^i$ to be independent. From the perspective of the dynamics on $\partial\Sigma$ this can be interpreted just as a (partial) gauge fixing condition on the, otherwise arbitrary, values of $\rho_\partial^i$. In any case it is important to remember that this condition comes from the requirement that the bulk and boundary dynamics must be consistent. Finally, it is clear that in the case $2\alpha_3\alpha_4-\alpha_1\alpha_5=0$ we recover standard GR with cosmological constant  \cite{Carlip} at the boundary with the same gauge fixing.

\subsection{Solving the constraints}\label{solving}

The preceding discussion shows that we have a detailed understanding of the dynamics of this model, in particular, the form of the Hamiltonian vector fields, the constraints and the physical interpretation. It is important, however, to find suitable parametrizations for the constraint submanifold as this is necessary to get consistent initial data for the evolution equations. Although we will not pursue this issue to its ultimate conclusion here, we feel that it provides a rationale for looking at the Hamiltonian formulation of the model (not directly connected with its eventual quantization) and shows the appropriateness and convenience of the geometric approach that we are following.

A first issue that we want to discuss (relevant, in particular, to check the consistency of the constraint algorithm) is the compatibility of the secondary constraints \eqref{c4} and \eqref{c5}. It is a simple exercise to show that \eqref{c3} are equivalent to the conditions
\begin{subequations}\label{conds}
\begin{align}
  F^i & =f^{ij}\epsilon_{jkl}e^k\wedge e^\ell\,, \\
  De^i & =t^{ij}\epsilon_{jkl}e^k\wedge e^\ell\,,
\end{align}
\end{subequations}
with both $f^{ij}$ and $t^{ij}$ symmetric in the $i,j$ indices. Although these conditions are obtained from constraints defined in the bulk, they extend to the boundary by continuity. Notice that, in order to fully characterize the solutions to the constraints \eqref{c1}-\eqref{c5}, it is still necessary to solve them for $A^i$ and $e^i$ in terms of the $f^{ij}$ and $t^{ij}$. These auxiliary objects will have then to satisfy some integrability conditions coming, for instance, from the Bianchi identity $DF^i=0$ and $DDe^i=0$ (on the constraint hypersurface). As mentioned in section \ref{sec_2connHK}, the fact that the field equations have solutions guarantees that these conditions can be met, however, the full characterization  of the solutions to the constraints requires additional work.

Let us consider now the constraints \eqref{c4}. In terms of the pullbacks of the fields to the boundary of $\Sigma$ they become
\begin{subequations}\label{condsboundary}
\begin{align}
  \big(2\alpha_2t^{mi}_\partial+\alpha_5f^{mi}_\partial+\alpha_4\delta^{mi}\big)\imath_\partial^*(\epsilon_{ijk}e^j\wedge e^k)=0\,, \\
  \big(\alpha_5t^{mi}_\partial+2\alpha_3f^{mi}_\partial+\alpha_1\delta^{mi}\big)\imath_\partial^*(\epsilon_{ijk}e^j\wedge e^k)=0\,.
\end{align}
\end{subequations}
In the same way, we get for \eqref{c5}
\begin{subequations}\label{condsboundary2}
\begin{align}
  \big(2\alpha_2t^{mi}_\partial+\alpha_5f^{mi}_\partial+\alpha_4\delta^{mi}\big)\imath_\partial^*(\epsilon_{ijk}e^j_\perp e^k)=0\,, \\
  \big(\alpha_5t^{mi}_\partial+2\alpha_3f^{mi}_\partial+\alpha_1\delta^{mi}\big)\imath_\partial^*(\epsilon_{ijk}e^j_\perp e^k)=0\,.
\end{align}
\end{subequations}
Although the structure of both sets of constraints in terms of $f^{ij}_\partial$ and $t^{ij}_\partial$ is similar, \textit{they are not the same}. In particular, it is not true that all of them are simply equivalent to
\begin{subequations}\label{condsboundary3}
\begin{align}
  2\alpha_2t^{ij}_\partial+\alpha_5f^{ij}_\partial+\alpha_4\delta^{ij}=0\,, \\
  \alpha_5t^{ij}_\partial+2\alpha_3f^{ij}_\partial+\alpha_1\delta^{ij}=0\,.
\end{align}
\end{subequations}

A detailed analysis of \eqref{condsboundary} and \eqref{condsboundary2} shows that the constraint submanifold is divided into several dynamical sectors among which the one given by the condition \eqref{sector} is the most relevant for our purposes and the only one discussed in the paper. Additional perspective on the parametrization of the constraint submanifold can be gained by looking at \eqref{c3} from a different point of view. Let us consider
\begin{subequations}
\begin{align}
  \epsilon^i_{\,jk}De^j\wedge e^k&=0\,,\label{constr1}\\
  \epsilon^i_{\,jk}e^j\wedge F^k&=0\,,\label{constr2}
\end{align}
\end{subequations}
and try to find suitable parametrizations of the phase space submanifold defined by them (forgetting now about boundary conditions). Let us start by looking at \eqref{constr1}. By expanding $A_i=a_{ij}e^j$ we immediately get
\begin{equation}\label{a}
e_{[i}\wedge \mathrm{d} e_{j]}-{\bm \omega} a_{[ij]}=0\,,
\end{equation}
so that
\begin{equation}\label{aij}
a_{ij}=\alpha_{ij}-\frac{\mathrm{d}(e_i\wedge e_j)}{2{\bm\omega}}
\end{equation}
with $\alpha_{ij}=\alpha_{ji}$ but, otherwise, arbitrary. Plugging now $A_i=a_{ij}e^j$ into \eqref{constr2} and using \eqref{a} we find out that it becomes
\begin{equation}\label{vector}
  \epsilon_{ij}^{\,\,\,\,\,k}e^j\wedge \mathrm{d}(a_{k\ell}e^\ell)+\frac{1}{2}a_{ij}\epsilon^j_{\,\,k\ell}\mathrm{d}(e^k\wedge e^\ell)=0\,.
\end{equation}
In terms of $\alpha_{ij}$ this is an inhomogeneous, \emph{linear} PDE, a somewhat surprising fact owing to the quadratic term in the connection appearing in the curvature $F^i$. Notice, however, that the analogous equation in geometrodynamics is linear in the momenta canonically conjugate to the 3-metric so this is not completely unexpected.

In order to find a complete parametrization of the constraint submanifold, it is necessary to solve \eqref{vector} together with the boundary constraints. Although it is not inconceivable that a closed form solution can be found, it is probably convenient to impose some simplifying conditions to render them easier to solve. These conditions, in fact, can be used also to reduce or eliminate the arbitrariness in the Hamiltonian vector field \eqref{VectorField} originating in the presence of the arbitrary functions $A_\perp^i$ and $e_\perp^i$. In this capacity, they are \emph{gauge fixing} conditions.

The issue of finding effective gauge fixing in non-abelian theories such as the ones presented here is a delicate one (Gribov ambiguity, topological obstructions, etc.) so we will not consider it further here. In any case, the simple form of the constraints and, in particular, the linearity of \eqref{vector} offers hope that a manageable solution to this problem exists.

\section{Conclusions and comments}\label{sec_Conclusions}

We have discussed in the paper several generalizations of the Pontryagin and HK actions to spacetime manifolds with boundary. The corresponding field equations consist of bulk and boundary contributions. The latter are equations involving the pullback of the fields. In a sense, they are boundary conditions, but in the models considered here they have interesting interpretations. In fact, by appropriately choosing the coupling constants $\bm \alpha$ it is possible to get, for instance, the 3-dimensional Euclidean Einstein equations with an arbitrary cosmological constant.

An interesting observation involving the relationship between the dynamics in the bulk and at the boundary is the following. On one hand, we can consider the \textit{restriction} of the dynamics in the bulk to the boundary. As shown in the paper, the geometrical meaning of the bulk dynamics is quite easy to describe: on initial data satisfying the constraints it reduces to Lie dragging along an arbitrary vector field tangent to the boundary and local internal $SO(3)$ rotations. The fact that the solution to the constraints have to satisfy additional conditions on the boundary of $\Sigma$ does not change the fact that the dynamics of the bulk fields at the boundary (defined by continuity of the bulk dynamics) has the simple interpretation that we have just spelled. On the other hand, we have to consider the dynamics of the \textit{pulled-back fields}. As we have shown, this is given by 3d-gravitational theories (determined by the choice of the $\bm \alpha$ parameters). The compatibility of the bulk and boundary dynamics originating in the consistency of the Dirac algorithm shows that these 3-dimensional theories admit a simple interpretation in terms of the bulk dynamics.

The Hamiltonian description for the actions discussed in the paper has been obtained by relying on the geometric approach developed in \cite{Dirac_geom} to specifically deal with field theories defined in regions with boundaries. Although Hamiltonian methods are often used as a first step to quantization, we have taken advantage of them in a purely classical context to get several interesting results. First of all, we have been able to extend the usual interpretation of the HK model to the case where boundaries are present. Furthermore, we have shown that the specific choice of parameters $\alpha_1=\alpha_2$ is such that there are no field equations in the bulk and the theory can be fully described on the boundary (as the Mielke-Baekler model). This reproduces the relationship between the Pontryagin and Chern-Simons field theories.

From a purely technical perspective, it is important to mention that some of the constraints appearing in our model, specifically \eqref{c5}, have their origin in the consistency of the bulk and boundary dynamics. Although this interplay may be expected \textit{a priori}, to our knowledge this is the first instance where this phenomenon is shown in a concrete example.

As far as the meaning of the constraints \eqref{c5} is concerned, it is interesting to see the role played by the condition \eqref{sector} in the interpretation of the dynamics of the sector that it defines. From the point of view of the boundary dynamics, it can be understood as a partial gauge fixing condition for the 3-dimensional gravitational part, whereas from the point of view of the bulk it forces the vector field $\bm \rho$ defining the dynamical diffeomorphisms of the HK model to be tangent to the boundary as it should.

Several questions remain open, for instance:
\begin{itemize}
  \item An obvious way to recover Lorentzian 3-dimensional extensions of GR at the boundary is to replace the internal $SO(3)$ group by the Lorentz group $SO(1,2)$. This will change the dynamics in the bulk to a \emph{Lorentzian} HK model which should be studied in detail.
  \item  A related important issue is the integrability of the Hamiltonian vector fields, the character of the field equations (hyperbolic or elliptic) and the type of conditions that have to be used to completely specify their solutions.
  \item As the phase space of the models considered here is the same as the one used in the Ashtekar formulation for GR, one could attempt a quantization for these systems, defined in spacetimes with boundary, inspired in the one used to describe quantum black holes in Loop Quantum Gravity.
  \item To study the non-generic case in which the $e^i$ do not define a frame (and hence describe degenerate metrics). In such situation, we expect to have extra constraints in addition to the ones presented in the paper.
\end{itemize}

These will be considered in future work.

\acknowledgments

This work has been supported by the Spanish Ministerio de Ciencia Innovaci\'on y Uni\-ver\-si\-da\-des-Agencia Estatal de Investigaci\'on/FIS2017-84440-C2-2-P grant. Bogar D\'{\i}az is supported by the CONACYT  (M\'exico) postdoctoral research fellowship N${^{\underline{\rm{o}}}}\,$371778. Juan Margalef-Bentabol is supported by 2017SGR932 AGAUR/Generalitat de Catalunya, MTM2015-69135-P/FEDER, MTM2015-65715-P. He is also supported in part by the Eberly Research Funds of Penn State, by the NSF grant PHY-1806356, and by the Urania Stott fund of Pittsburgh foundation UN2017-92945.


\begin{thebibliography}{99}


\bibitem{Chern}  S-S. Chern  and  J. Simons, \emph{Characteristic Forms and Geometric Invariants},  \href{http://doi.org/10.2307/1971013}{\emph{Annals Math.} {\bf 99} (1974) 48}.

\bibitem{FB1}  J. F. Barbero G., \emph{General Relativity as a Theory of Two Connections},  \href{https://doi.org/10.1142/S0218271894000587}{\emph{Int. J. Mod. Phys. D} {\bf 3} (1994) 397}.

\bibitem{HK} V. Husain  and K. Kucha\v{r}, \emph{General covariance, new variables, and dynamics without dynamics}, \href{https://doi.org/10.1103/PhysRevD.42.4070}{\emph{Phys. Rev. D} {\bf 42} (1990) 4070}.

\bibitem{Mielke} E. W. Mielke and P. Baekler, \emph{Topological Gauge Model Of Gravity With Torsion}, \href{https://doi.org/10.1016/0375-9601(91)90715-K}{\emph{Phys. Lett. A}  {\bf 156} (1991) 399}.

\bibitem{MBH} P. Baekler, E. W. Mielke and F. W. Hehl, \emph{Dynamical symmetries in topological 3D gravity with torsion}, \href{https://doi.org/10.1007/BF02726888}{\emph{Nuovo Cim. B} {\bf 107} (1992) 91}.

\bibitem{Kawai} T. Kawai, \emph{Teleparallel theory of (2+1)-dimensional gravity}, \href{https://doi.org/10.1103/PhysRevD.48.5668}{\emph{Phys. Rev. D} {\bf 48} (1993)
5668}.

\bibitem{BV1} M. Blagojevi\'c and M. Vasili\'c, \emph{Asymptotic dynamics in 3D gravity with torsion}, \href{https://doi.org/10.1103/PhysRevD.68.124007}{\emph{Phys. Rev. D} {\bf 68} (2003) 124007}.

\bibitem{BV2} M. Blagojevi\'c and M. Vasili\'c, \emph{Asymptotic symmetries in 3D gravity with torsion}, \href{https://doi.org/10.1103/PhysRevD.67.084032}{\emph{Phys. Rev. D} {\bf 67} (2003) 084032}.

\bibitem{Cacciatori} S. L. Cacciatori, M. M. Caldarelli, A. Giacomini, D. Klemm and D. S. Mansi, \emph{Chern-Simons formulation of three-dimensional gravity with torsion and nonmetricity}, \href{https://doi.org/10.1016/j.geomphys.2006.01.006}{\emph{J. Geom. Phys.} {\bf 56} (2006) 2523}.

\bibitem{BGMR} R. Banerjee, S. Gangopadhyay, P. Mukherjee and D. Roy, \emph{Symmetries of the general topologically
massive gravity with torsion in the Hamiltonian and Lagrangian formalisms}, \href{https://doi.org/10.1007/JHEP02(2010)075}{\emph{JHEP} {\bf 02} (2010) 075}.

\bibitem{BC} M. Blagojevi\'c and B. Cvetkovi\'c, \emph{Conserved charges in 3D gravity}, \href{https://doi.org/10.1103/PhysRevD.81.124024}{\emph{Phys. Rev. D} {\bf 81} (2010) 124024}.

\bibitem{Livine} V. Bonzom and E. R. Livine, \emph{A Immirzi-like parameter for 3D quantum gravity}, \href{https://doi.org/10.1088/0264-9381/25/19/195024}{\emph{Class. Quantum Grav.} {\bf 25} (2008) 195024}.

\bibitem{Basu} R. Basu and S. K. Paul, \emph{2+1 quantum gravity with a Barbero-Immirzi-like parameter on toric spatial foliation}, \href{https://doi.org/10.1088/0264-9381/27/12/125003}{\emph{Class. Quantum Grav.} {\bf 27} (2010) 125003}.

\bibitem{Dirac_geom} J. F. Barbero G., B. D\'{i}az, J. Margalef-Bentabol and E. J. S. Villase\~{n}or, \emph{Dirac's algorithm in the presence of boundaries: a practical guide to a geometric approach}, accepted for publication in Class. Quantum Grav. \href{https://arxiv.org/abs/1904.11790}{arXiv:1904.11790}.

\bibitem{Gotay2} M. Gotay, J. Nester and G. Hinds, \emph{Presymplectic manifolds and the Dirac--Bergmann theory of constraints}, \href{https://doi.org/10.1063/1.523597}{\emph{J. Math. Phys.} \textbf{19} (1978) 2388}.

\bibitem{BPV} J. F. Barbero G., J. Prieto and E. J. S. Villase\~nor, \emph{Hamiltonian treatment of linear field theories in the presence of boundaries: a geometric approach},  \href{https://doi.org/10.1088/0264-9381/31/4/045021}{\emph{Class. Quantum Grav.} \textbf{31} (2014) 045021}.

\bibitem{tesisJuan} J. Margalef-Bentabol,  \emph{Towards general relativity through parametrized theories}, Doctoral thesis, \href{https://arxiv.org/abs/1807.05534}{arXiv:1807.05534}.


\bibitem{barbero} J. F. Barbero G., \emph{Reality conditions and Ashtekar variables: A different perspective}, \href{https://doi.org/10.1103/PhysRevD.51.5498}{\emph{Phys. Rev. D} {\bf 51} (1995) 5498}.

\bibitem{Carlip} S. Carlip, \emph{Quantum Gravity in 2+1 dimensions}, Cambridge University Press (2003).

% Please avoid comments such as "For a review'', "For some examples",
% "and references therein" or move them in the text. In general,
% please leave only references in the bibliography and move all
% accessory text in footnotes.

% Also, please have only one work for each \bibitem.


\end{thebibliography}
\end{document}